\documentstyle[eqsecnum,preprint,aps,prd]{revtex}
\def\be{\begin{equation}}
\def\ee{\end{equation}}
\def\pa{\partial}
\def\eps{\epsilon}

\def\rarrow{\rightarrow}

\begin{document}
\draft
\tighten
\title{Self dual models and mass generation in planar field theory}
\author{Rabin Banerjee {\footnote {rabin@boson.bose.res.in}} and 
        Sarmishtha Kumar {\footnote {kumar@boson.bose.res.in}}}
\address{
S.N. Bose National Centre For Basic Sciences\\
Salt Lake City, Block JD, Sector III\\
Calcutta 700 098, INDIA\\ }
\maketitle

\begin{abstract} {We analyse in three space-time dimensions, the connection between abelian self 
dual vector doublets and their counterparts containing both an explicit mass and a topological mass. Their correspondence is established in the lagrangian formalism using an operator approach as 
well as a path integral approach. A canonical hamiltonian analysis is presented, which also
shows the equivalence with the lagrangian formalism.
The implications of our results for bosonisation in three dimensions are 
discussed.}
\end{abstract}
\section{Introduction}
Self dual models in three space time dimensions, have certain distinct features which are 
essentially connected with the presence of the Chern-Simons term which is both metric and gauge independent. An important variant of such a model is the topologically massive gauge theory
\cite{S,DJT} where gauge invariance coexists with the finite mass, single helicity and parity
violating nature of the excitations. Its dynamics is governed by a lagrangian comprising
both the Maxwell and Chern-Simons terms. The equations of motion, when expressed in terms of
the dual to the field tensor, manifest a self duality. An equivalent version of this model
also exists, where the self duality is revealed in the equations of motion for the basic field
\cite{TPN,DJ,BR}. More recently, another possibility has been considered where, instead of 
the first derivative Chern-Simons term, a parity violating third derivative term is added
to the Maxwell term \cite{DJ1}.

An intriguing fact, first noted in \cite{DJT} and briefly discussed in \cite{D,BW,BK},
is that topologically massive doublets, with identical mass parameters having opposite sign,
are equivalent to a parity preserving vector theory with an explicit mass term. This is the Proca model. The invariance of the doublets under the combined parity and field interchanges is thereby easily
understood from  the equivalent theory. Moreover the two theories of the doublet characterise
self and anti-self dual solutions, depending on the sign of the mass term.
 The final effective theory, which is a superposition of these solutions, 
therefore hides these symmetries.

In this paper we will make a detailed analysis of a doublet of topologically massive theories
with distinct mass parameters. The resultant theory is a parity violating non-gauge vector
theory with explicit as well as topological mass terms. This is demonstrated in section-II
in the lagrangian formalism using an operator approach. These results are then interpreted in
 the path integral approach. A hamiltonian reduction of the effective theory, based on canonical 
transformations, is performed in section-III. The diagonalisation of the hamiltonian
reveals the presence of two massive modes, which are a combination of 
topological and explicit mass parameters. These modes can be identified with 
those of the original Maxwell-Chern-Simons doublet thereby revealing a complete equivalence with the lagrangian formalism. 
The diagonalisation of the energy-momentum tensor is carried
out in section-IV. Following a method elaborated in \cite{DJT}, the spin of 
the excitations is calculated. The helicity states are $\pm 1$, corresponding 
to the two modes of the theory. An application to the
bosonisation of a doublet of massive Thirring models in the long wavelength 
limit is discussed in section-V. Our concluding remarks are left for  
section-VI.

\section{Lagrangian analysis}
\hskip -.4 cm {\large{\bf{II.I An operator approach}}}
\vskip .5 cm
In this section we shall consider a doublet of self and anti-self dual models 
whose dynamics is governed, respectively, by the following lagrangian densities,\\
\be
{\cal L}_{SD} = {\cal L} _{-} = \frac{m_{-}}{2} g_{\mu} g^{\mu} - \frac{1}{2} 
\eps_{\mu \nu\lambda} g^{\mu} \pa ^{\nu} g ^{\lambda}\label{a} 
\ee
\be
{\cal L}_{ASD} = {\cal L} _{+} = \frac{m_{+}}{2} f_{\mu} f^{\mu} + \frac{1}{2}
\eps_{\mu \nu\lambda} f^{\mu} \pa ^{\nu} f ^{\lambda}\label{b}
\ee

The property of self (or anti-self) duality follows on exploiting the equations
of motion \cite{BW}. Note that the mass parameters are different in the two cases. 
It has been suggested \cite{BK} that the above models combine to yield the Maxwell-Chern- Simons model
with a conventional mass term. Here we quickly review that approach, 
which is based on \cite{MS}.  The idea is to construct an
effective lagrangian that will characterise the doublet.
Obviously a simple minded addition of the two lagrangians will not yield
anything. A new field will have to be introduced which will glue or solder
the two lagrangians. The final or effective lagrangian will not contain
this new field. Later on we shall show in what sense this approach can be
understood as an ``addition" of the two lagrangians. Consider the variation
of the lagrangians under the local transformation,

\be
\delta f_{\mu}=\delta g_{\mu}=\Lambda_{\mu}(x)
\label{1}
\ee
\\
The requisite variations are given by,

\be
\delta{\cal L}_{\mp} = J_{\mp}^{\mu}\Lambda_{\mu}
\label{2}
\ee
\\
where the currents are defined as,

\be
J^{\mu}_{\mp} = m_{\mp}h^{\mu} \mp \eps^{\mu\alpha\beta}\pa_{\alpha}h_{\beta}
 \,  ;\,\,\,  h = f,g
\label{3}
\ee
\\
Next we introduce the soldering field $W_\mu$ transformimg as,

\be
\delta W_{\mu} = \Lambda _{\mu}
\label{4}
\ee
\\
It is now simple to check that the following lagrangian,

\be
{\cal L}={\cal L}_{-}(g)+{\cal L}_{+}(f) - W_{\mu}(J_+^{\mu}(f) +
J_-^{\mu}(g)) + \frac{1}{2}(m_+ + m_-)W_{\mu}W^{\mu}
\label{5}
\ee
\\
is invariant under the transformations introduced earlier. The field 
$W_\mu$ plays the role of an auxiliary variable that can be eliminated by
using the equation of motion,

\be
W_{\mu} = \frac{1}{m_+ + m_-}(J_{\mu}^+(f) + J_{\mu}^-(g))
\label{6}
\ee

The final theory is manifestly invariant under the transformations
containing only the difference of the original fields. It is given by,

\be
{\cal L} = -\frac{1}{4} F_{\mu\nu}F^{\mu\nu}(A) + \frac{1}{2} \eps_{\mu \nu\lambda}
(m_{-} - m_{+})A^{\mu}\pa^{\nu}A^{\lambda} + \frac{1}{2} m_{+}m_{-}A_{\mu}A^{\mu} 
\label{j}
\ee   
where,\\
\be
A_\mu = \frac{1}{\sqrt {m_{+} + m_{-}}} (f_\mu - g_\mu ) \label{k}
\ee 
This is the Maxwell-Chern-Simons theory with an explicit mass term. A word
about the degree of freedom count might be useful. The lagrangians
(\ref{a}) and (\ref{b}) individually correspond to single massive modes. The
composite model (\ref{j}) corresponds to two massive modes. There is thus
a matching of the degree of freedom count.

It is now possible to take a different variation
of the fields, but the final result will be the same. To illustrate this 
consider, instead of (\ref{1}), the following variations,\\ 
\be
\delta f_{\mu} = \delta g_{\mu} = \eps_ {\mu\alpha\beta} \pa^{\alpha}\Lambda^{\beta}
\label{c}
\ee
The variations in the individual lagrangians can be written in terms of the parameter $ \Lambda$ as,\\
\be
\delta{\cal L}_{\mp} = J_{\mp}^{\alpha\beta} \pa_{\alpha}\Lambda_{\beta}\label{d}
\ee
where,\\
\be
J_{\mp}^{\alpha\beta} = m_{\mp}\eps^{\alpha\beta\mu} h_{\mu} \hskip .1 cm {\mp}\hskip .1 cm 
h^{\alpha\beta}; \hskip .25 cm h = f, g \label{e}
\ee
and,\\
\be
h^{\alpha\beta} = \pa^{\alpha} h^{\beta} - \pa^{\beta} h^{\alpha} \label{f}
\ee
Introducing an antisymmetric tensor field $ B_{\alpha\beta} $ transforming as,\\
\be
\delta B_{\alpha\beta} = \pa_{\alpha} \Lambda_{\beta} - \pa_{\beta}\Lambda_{\alpha}
\label{g}
\ee
it is possible to write a modified lagrangian,\\
\be
{\cal L} = {\cal L}_{SD} + {\cal L}_{ASD} - \frac{1}{2} B^{\alpha\beta}
({J_{\alpha\beta}^{+}(f) + J_{\alpha\beta}^{-}(g)}) + \frac{1}{4}
(m_{+} + m_{-}) B_{\alpha\beta}B^{\alpha\beta} \label{h}
\ee
that is invariant under (\ref{c}) and (\ref{g}); i.e. $\delta{\cal L} =  0 $.
Since $B_{\alpha\beta}$ is an auxiliary field it is eliminated from (\ref{h})
by using its solution. The final effective theory is just (\ref{j}).

The above manipulations have shown that it is possible to glue the two
lagrangians by introducing an auxiliary variable. We could adopt this method
to glue any two lagrangians; however the final result would not be local.
The local expression follows precisely because the self and anti-self dual
nature of the lagrangians engage in a cancelling act. Note that the variations
considered here lead to the combination $f_\mu -g_\mu$ in the effective theory.
By considering the variations with opposite signatures we would have been led
to the same effective theory but with the combination $f_\mu +g_\mu$.

As announced earlier we now show how the above approach enables one to 
directly obtain the effective theory by adding the two lagrangians,

\be
{\cal L}={\cal L}_+(f) + {\cal L}_-(g)
\label{7}
\ee

Introducing the combination (\ref{k}), we find,

\begin{eqnarray}
{\cal L} &=& {\cal L}_+(\sqrt{m_+ + m_-}A + g) + {\cal L}_-(g)
\nonumber\\
& = &\frac{m_+}{2}(m_+ + m_-)A^{\mu}A_{\mu} + \frac{1}{2}(m_+ +m_-)
g^{\mu}g_{\mu} + \sqrt{m_++m_-}\eps_{\mu \nu \lambda}g^{\mu}\pa^{\nu}
A^{\lambda}  \nonumber\\&+&  m_+\sqrt{m_++m_-}A_{\mu}g^{\mu}  
+ \frac{m_++m_-}{2}\eps_{\mu \nu\lambda}A^{\mu}\pa^{\nu}A^{\lambda}
\label{8}
\end{eqnarray}
\\
Now $g_\mu$ behaves as an auxiliary variable. It is eliminated in favour
of the other variable by using
the equation of motion. The end result reproduces (\ref{j}).

The compatibility of the equations of motion of the doublet and the effective 
theory is next shown. From (\ref{a}) and (\ref{b}) the following equation are
obtained,
\be
g_{\mu} = \frac{1}{m_{-}} \eps_{\mu\nu\lambda} \pa^{\nu}g^{\lambda} 
\label{a1}
\ee
\be 
\pa_{\beta} g^{\mu\beta} =  m_{-}\eps^{\mu\alpha\beta}\pa_{\alpha}g_{\beta}
\label{a2}
\ee
and,
\be
f_{\mu} = - \frac{1}{m_{+}}\eps_{\mu\nu\lambda}\pa^{\nu}f^{\lambda}\label{a3}
\ee
\be
\pa_{\beta}f^{\mu\beta} = - m_{+}\eps^{\mu\alpha\beta}\pa_{\alpha}f_{\beta}
\label{a4}
\ee
Using the above sets of equations it follows that,
\be
- \pa^{\nu} (f_{\mu\nu} - g_{\mu\nu}) + (m_{-} - m_{+}) \eps_{\mu\nu\lambda}
\pa^{\nu} (f^{\lambda} - g^{\lambda}) + m_{+}m_{-}(f_{\mu} - g_{\mu}) = 0
\label{a5}
\ee
which is just the equation of motion for the effective lagrangian (\ref{j}) 
with the identification (\ref{k}).

Now the self dual model is known to be equivalent to the Maxwell-Chern-Simons
theory \cite{DJ,BR}. Consequently the above analysis can be repeated for
a doublet of Maxwell-Chern-Simons theories defined by the lagrangian 
densities,\\
\be
{\cal L}_{-}(P) = - \frac{1}{4m_{-}}F_{\mu\nu}F^{\mu\nu}(P) + \frac{1}{2}
\eps_{\mu\nu\lambda}P^{\mu}\pa^{\nu}P^{\lambda} 
\label{l} 
\ee
\be
{\cal L}_{+}(Q) = - \frac{1}{4m_{+}}F_{\mu\nu}F^{\mu\nu}(Q) - \frac{1}{2}
\eps_{\mu\nu\lambda}Q^{\mu}\pa^{\nu}Q^{\lambda} 
\label{m}  
\ee
Specifically, the models (\ref{l}) and (\ref{m}) are the analogues of those 
given in (\ref{a}) and (\ref{b}), respectively. For the sake of comparison,
the mass parameters $ m_{\mp} $ are taken to be identical in both cases.  

Now consider the variations of the lagrangians under the following transformations,
\be
\delta P_{\mu} = \delta Q_{\mu} = \Lambda_{\mu} \label{n}
\ee
Then it follows,\\
\be
\delta{\cal L}_{\mp} = J_{\mu\nu}^{\mp} \pa^{\mu}\Lambda^{\nu} \label{o}
\ee
where,\\
\be
J_{\mu\nu}^{\mp}(W) = - \frac{1}{m_{\mp}} F_{\mu\nu}(W)  \pm  \eps_{\mu\nu\lambda} 
W^{\lambda};\hskip .25 cm W = P, Q \label{p} 
\ee
Introducing the $B_{\mu\nu}$ field transforming as (\ref{g}), it is seen that
the following combination,\\
\be
{\cal L} = {\cal L}_{-}(P) + {\cal L}_{+}(Q) - \frac{1}{2} B_{\mu\nu}
( J_{+}^{\mu\nu} + J_{-}^{\mu\nu} ) - \frac{1}{4} ( \frac{1}{m_{+}} + 
\frac{1}{m_{-}})B_{\mu\nu}B^{\mu\nu} \label{q} 
\ee
is invariant under the relevant transformations (\ref{g}) and (\ref{n}).

As before, the auxiliary field $B_{\mu\nu}$ is eliminated from (\ref{q}) to yield
the lagrangian (\ref{j}) in terms of a composite field which is the difference of
the fields in the doublet,\\
\be
A_\mu = \frac{1}{\sqrt {m_{+} + m_{-}}} (P_\mu - Q_\mu ) \label{r}
\ee

The other considerations discussed for the self-dual models are all 
applicable here.

\vskip .5 cm
\hskip -.4 cm{\large{\bf{ II.II  Path integral derivation}}}
\vskip .3 cm
The above discussion has a natural interpretation 
in the path integral formalism.
The point is that the analysis related to equations (\ref{7}) and (\ref{8})
shows that it is possible to obtain the effective theory by an addition of the
lagrangians and then identifying an auxiliary variable which is eventually
eliminated. Since the problem is gaussian it is straightforward to interpret
it in the path integral language. 
The elimination of the auxiliary variable just corresponds to a gaussian
integration over that variable. 
Let us therefore consider the following generating functional 
{\footnote{Note that the path integral following from the hamiltonian version 
\cite{BR} requires the factor $\delta (f_{0} + \frac{1}{m_{+}} \eps_{ij}\pa_{i}
f_{j} ) 
\delta 
(g_{0} - \frac{1}{m_{-}}\eps_{ij}\pa_{i}g_{j}) $ in the measure to account for
the constraints. Since this is a Gaussian problem the result of the path
integral remains unaltered even if these factors are not included. This is how
we choose to define the basic lagrangian path integral for the self and anti
self dual models.}} 
for the doublet of self and anti-self dual models (\ref{a}) and (\ref{b}),\\
\begin{eqnarray}
{\cal Z } & = & \int df_{\mu} dg_{\mu}\exp i\int d^{3}x [{\cal L}_{-}(g) + {\cal L}_{+}
(f) \nonumber \\
& & + \frac{1}{\sqrt {m_{+} + m_{-}}}(f_{\mu} - g_{\mu}) J^{\mu}]
\end{eqnarray} \label{s}
where a source has been introduced that is coupled to the difference (\ref{k}) of
the variables. A relabeling of variables as in (\ref{k}) is made for which the 
jacobian is trivial. The path integral is now rewritten in terms of the redefined
variable $A_{\mu}$ and $g_{\mu}$,\\

\begin{eqnarray}
{\cal Z } & = & \int dA_{\mu}  dg_{\mu} \exp i \int d^{3}x [ \frac{m_{+}}{2}(\sqrt 
{m_{+} + m_{-}} A_{\mu} + g_{\mu} )^2 \nonumber\\
& & + \frac{1}{2} \eps_{\mu\nu\lambda} (\sqrt {m_{+} + m_{-}} A^{\mu} + g^{\mu} )
\pa^{\nu} (\sqrt {m_{+} + m_{-}}A^{\lambda} + g^{\lambda} ) \nonumber\\ 
& & + \frac{m_{-}}{2} g_{\mu}g^{\mu}- \frac{1}{2} \eps_{\mu\nu\lambda} g^{\mu}
\pa^{\nu}g^{\lambda} + A_{\mu}J^{\mu}]
\end{eqnarray}\label{t}   
Integrating over the $g_{\mu}$ variable yields,
\begin{eqnarray}
{\cal Z } &  = & \int dA_{\mu} \exp i \int d^{3}x [ - \frac{1}{4} F_{\mu\nu}F^{\mu\nu}
 + \frac{1}{2} (m _{-} - m_ {+} ) \eps_{\mu\nu\lambda}A^{\mu}\pa^{\nu}
A^{\lambda}\nonumber \\
& & + \frac{m_{+}m_{-}}{2} A_{\mu}A^{\mu} + A^{\mu} J_{\mu}]\label{u}
\end{eqnarray}
In the absence of sources this is just the partition function for the 
Maxwell-Chern-Simons-Proca model (\ref{j}). Furthermore, the $ A_{\mu} $ 
field in
(\ref{u}) is related to the original doublet fields by exactly the same equation
(\ref{k}). This shows the equivalence of the results obtained by the two approaches.

It is equally possible to carry out a similar analysis for a doublet of Maxwell-
Chern-Simons theories. However, a gauge fixing is necessary to account for the gauge
invariance of these theories. As was shown in \cite{BR}, through the use of master 
lagrangians, the basic field in the self dual model can be identified with the basic 
field in the Maxwell-Chern-Simons theory defined in the covariant gauge. We therefore
consider the generating functional obtained from (\ref{l}),(\ref{m});  
\begin{eqnarray} 
{\cal Z } & = & \int dP_{\mu} dQ_{\mu} \delta(\pa_{\mu}P^{\mu}) \delta(\pa_
{\mu}Q^ {\mu}) \exp i \int d^{3}x [{\cal L}_{-}(P) + {\cal L}_{+}(Q) 
\nonumber \\
&  & \hskip 2 cm  + \frac{1}{\sqrt {m_{+} + m_{-}}}(P_{\mu} - Q_{\mu})  J^{\mu}]
\end{eqnarray} \label{v} 
where, as before, a coupling with an external source has been done with the difference
(\ref{r}) of the variables. Because of the gauge invariance of the integrand, 
the source $ J_{\mu} $ 
should be conserved. 

To perform the path integration, a renaming of variables according to (\ref{r}) is done for which
the jacobian is trivial. Then, 
\begin{eqnarray}
{\cal Z } & = & \int dA_{\mu} dQ_{\mu} \delta(\pa_{\mu}A^{\mu}) \delta(\pa_{\mu}
Q^ {\mu}) \exp i \int d^{3}x [ - \frac{1}{4m_{-}}(m_{+} + m_{-}) F_{\mu\nu}(A)
F^{\mu\nu}(A) \nonumber\\
& & - \frac{1}{4}(\frac{1}{m_{+}} + \frac{1}{m_{-}})F_{\mu\nu}(Q)F^{\mu\nu}(Q)
- \frac{\sqrt{m_{+} + m_{-}}}{2m_{-}}F_{\mu\nu}(A)F^{\mu\nu}(Q) \nonumber\\
& & + \sqrt{m_{+} + m_{-}} \eps_{\mu\nu\lambda} Q^{\mu}\pa^{\nu}A^{\lambda}
 + \frac{1}{2}(m_{+} + m_{-})\eps_{\mu\nu\lambda} A^{\mu}\pa^{\nu}A^{\lambda}
 + A_{\mu}J^{\mu}]
\end{eqnarray} \label{w}
Performing the integral over the $ Q_{\mu}$ variables yields,
\begin{eqnarray}
{\cal Z } & = & \int dA_{\mu} \delta(\pa_{\mu}A^{\mu}) \exp i \int d^{3}x 
[- \frac{1}{4} F_{\mu\nu}(A)F^{\mu\nu}(A) + \frac{1}{2}(m_{+}m_{-}) 
A_{\mu}A^{\mu} \nonumber\\
& & + \frac{1}{2}(m_{-} - m_{+})\eps_{\mu\nu\lambda} A^{\mu}\pa^{\nu}A^{\lambda}
 + A_{\mu}J^{\mu}]
\end{eqnarray}\label{x}
Express the delta function in the measure by an integral over a variable 
$ \alpha$,
\begin{eqnarray}
{\cal Z } & = & \int dA_{\mu} d \alpha \exp i \int d^{3}x [\alpha \pa_{\mu}
A^{\mu} - \frac{1}{4} F_{\mu\nu}(A)F^{\mu\nu}(A) + \frac{1}{2}(m_{+}m_{-})
A_{\mu}A^{\mu} \nonumber\\
& & + \frac{1}{2}(m_{-} - m_{+})\eps_{\mu\nu\lambda} A^{\mu}\pa^{\nu}A^{\lambda}
 + A_{\mu}J^{\mu}]
\end{eqnarray}\label{y}
Introducing a St\"{u}ckelberg transformed field $ A_{\mu} \rightarrow A_{\mu} + 
(m_+m_-)^{-1}
\pa_{\mu}\alpha $ and using the conservation of the source $( i.e. \pa_{\mu} J^{\mu} = 0) $
it follows that, 
\begin{eqnarray}
{\cal Z } & = & \int dA_{\mu}\exp i \int d^{3}x [- \frac{1}{4} F_{\mu\nu}F^{\mu\nu}
 + \frac{1}{2}(m_{+} m_{-})A_{\mu}A^{\mu}\nonumber\\
& &  + \frac{1}{2}(m_{-} - m_{+})\eps_{\mu\nu\lambda} A^{\mu} \pa^{\nu}A^{\lambda} 
+ A_{\mu}J^{\mu}] \label{z}
\end{eqnarray}
where the integral over $ \alpha$ has been absorbed in the normalisation.

As before, the generating functional for the Maxwell-Chern-Simons theory with an explicit mass term is obtained. The connection of the basic field $A_{\mu}$ 
with the original doublet,of course, remains the same as in (\ref{r}).

\section{Hamiltonian reduction and canonical transformations} 
The results of the previous section were achieved in the lagrangian formulation
by combining the doublet to yield the composite model. A complementary viewpoint
will now be presented in the hamiltonian formulation. By solving the constraint,
the hamiltonian of the model is expressed in term of a reduced set of variables.
Next, by means of a canonical transformation, the hamiltonian gets decomposed
into two distinct pieces, which correspond to the hamiltonians of the Maxwell-
Chern-Simons doublet. This technique of using canonical transformations to
diagonalise a hamiltonian is of course well known and appears in defferent
versions and different situations. More recently, in the context of the
lagrangian formalism discussed in section II.I, it has been developed
in \cite{AW}.  Defining a new set of parameters,
\begin{eqnarray}
m_{-} - m_{+} & =  & \theta \nonumber\\
m_{+}m_{-} & = & m^{2} \label{aa} 
\end{eqnarray} 
the lagrangian (\ref{j}) takes the form,
\be
{\cal L } = - \frac{1}{4} F_{\mu\nu}F^{\mu\nu} + \frac{\theta}{2} \eps_
{\mu\nu\lambda} A^{\mu} \pa^{\nu}A^{\lambda} + \frac{m^{2}}{2} A_{\mu}A^{\mu} 
\label{ab}
\ee 
The canonical momenta are,
\be 
\pi_{i} = \frac{\pa{\cal L}}{\pa \dot{A}^{i}} = - (F_{0i} + \frac{\theta}{2} 
\eps_{ij} A_{j}) \label{ac}
\ee 
while,
\be
\pi_{0} \approx 0 \label{ad} 
\ee 
is the primary constraint. The canonical hamiltonian is given by,
\be
H = \frac{1}{2} \int d^2x [\pi_{i}^{2} + \frac{1}{2}F_{ij}^{2} + (\frac{\theta^
{2}}{4} + m^{2})A_{i}^{2} - \theta  \eps_{ij}A_{i}\pi_{j} + m^{2}A_{0}^{2}]
 + \int d^2x A_{0} \Omega  \label{ae}
\ee  
where,
\be
\Omega = \pa_{i}\pi_{i} - \frac{\theta}{2} \eps_{ij}\pa_{i}A_{j} - m^{2} A_{0}
\approx 0 \label{af} 
\ee 
is the secondary constraint. Eliminating the multiplier $A_{0}$ from (\ref{ae})
by solving the constraint (\ref{af}) we obtain,
\begin{eqnarray}
H & = & \frac{1}{2}\int d^2x [\pi_{i}^{2} + (\frac{1}{2} + \frac{\theta^{2}}
{8m^{2}}) F_{ij}^{2} + ( \frac{\theta^{2}}{4} + m^{2}) A_{i}^{2} - \theta
\eps_{ij}A_{i}\pi_{j}] \nonumber \\ 
& & + \frac{1}{2m^{2}} \int d^2x [( \pa_{i}\pi_{i})^{2} - \theta \pa_{i}\pi_{i}
\eps_{lm}\pa_{l}A_{m}] \label{ag} 
\end{eqnarray} 
Making the canonical transformations in terms of the new canonical pairs
$ (\alpha, \pi_{\alpha}) $ and $ (\beta, \pi_{\beta})$,
\begin{eqnarray}
A_{i} & = & \frac{2m}{\sqrt{4m^{2} + \theta^{2}}} \eps_{ij}\frac{\pa_{j}}{\sqrt
{-\pa^{2}}} (\alpha + \beta) + \frac{1}{2m} \frac{\pa_{i}}{\sqrt{-\pa^{2}}}
(\pi_{\alpha} - \pi_{\beta})\nonumber \\
\pi_{i} & = & - \frac{\sqrt{4m^{2} + \theta^{2}}}{4m} \eps_{ij}\frac{\pa_{j}}
{\sqrt{-\pa^{2}}}(\pi_{\alpha} + \pi_{\beta}) + m \frac{\pa_{i}}{\sqrt{-\pa^
{2}}} (\alpha - \beta) \label{ah}
\end{eqnarray}
the hamiltonian decouples into two independent pieces,
\be
H (A_{i}, \pi_{i}) = H (\alpha, \pi_{\alpha}) + H (\beta, \pi_{\beta})
\ee \label{ai}
where,

\begin{eqnarray}
 H (\alpha, \pi_{\alpha}) & = & \frac{1}{16m^{2}} \sqrt{4m^{2} + \theta^{2}}
(\sqrt{4m^{2} + \theta^{2}} - \theta)\int d^{2}x \pi_{\alpha}^{2} + 
 \frac{\sqrt{4m^{2} + \theta^{2}} + \theta}{\sqrt{4m^{2} + \theta^{2}}} \int 
d^{2}x(\pa_{i}\alpha )^{2} \nonumber\\ 
& & \hskip 1cm  + m^{2}\frac{(\sqrt{4m^{2} + \theta^{2}} - \theta)}{\sqrt
{4m^{2} + \theta^{2}}} \int d^{2}x \alpha^{2} \nonumber \\
H (\beta, \pi_{\beta}) & = & \frac{1}{16m^{2}} \sqrt{4m^{2} + \theta^{2}}
(\sqrt{4m^{2} + \theta^{2}} + \theta)\int d^{2}x \pi_{\beta}^{2} +
 \frac{\sqrt{4m^{2} + \theta^{2}} - \theta}{\sqrt{4m^{2} + \theta^{2}}}
\int d^{2}x(\pa_{i}\beta )^{2} \nonumber \\
& & \hskip 1cm  + m^{2}\frac{(\sqrt{4m^{2} + \theta^{2}} + \theta)}
{\sqrt{4m^{2} + \theta^{2}}} \int d^{2}x \beta^{2} 
\end{eqnarray} \label{aj}

To recast these expressions in a familiar form, a trivial scaling is done,
\begin{eqnarray}
\alpha^{2} &\rightarrow &\frac{1}{2}\frac{\sqrt{4m^{2} + \theta^{2}}}{\sqrt{4m^
{2} + \theta^{2}} + \theta} \alpha^{2},\hskip .2 cm 
{\pi_{\alpha}^{2}}  \rightarrow 2 \frac{\sqrt{4m^{2} + \theta^{2}} + \theta}
{\sqrt{4m^{2} + \theta^{2}}} \pi_{\alpha}^{2} \nonumber\\
\beta^{2} &\rightarrow &\frac{1}{2}\frac{\sqrt{4m^{2} + \theta^{2}}}{\sqrt{4m^
{2} + \theta^{2}} - \theta} \beta^{2},\hskip .2 cm \pi_{\beta}^{2} \rightarrow 
2 \frac{\sqrt{4m^{2} + \theta^{2}} - \theta}{\sqrt{4m^{2} + \theta^{2}}}
\pi_{\beta}^{2} \label{ak}
\end{eqnarray}
so that,
\begin{eqnarray}
H (\alpha, \pi_{\alpha}) & = & \frac{1}{2}\int d^{2}x [(\pa_{i}\alpha)^{2} +
 \pi_{\alpha}^{2} + m_{+}^{2}\alpha^{2}] \nonumber\\
H (\beta, \pi_{\beta}) & = &\frac{1}{2}\int d^{2}x [(\pa_{i}\beta)^{2} +
\pi_{\beta}^{2}+ m_{-}^{2}\beta^{2}] \label{al}
\end{eqnarray}
with, 
\be
m_{\pm} = \sqrt{m^{2} + \frac{\theta^{2}}{4}} \mp  \frac{\theta}{2}\label{am}
\ee 
These relations show that the theory possesses two massive modes with mass
$m_{+}$ and $m_{-}$ which satisfy the Klein Gordon equation.
Furthermore since $ m_{\pm}$ in (\ref{am}) are the solutions to the set 
(\ref{aa}), these can be identified with the corresponding mass parameters 
occurring in the Maxwell-Chern-Simons doublet (\ref{l}) and (\ref{m}).
The above hamiltonians are indeed the reduced expressions obtained
from (\ref{m}) and (\ref{l}), respectively. The canonical reduction of the 
Maxwell-Chern-Simons theory has been done in \cite{DJT} but we present it 
here from our viewpoint for the sake of completeness. Let us, for instance, 
consider the lagrangian (\ref{l})\footnote{The variable P, for convenience,
is now called A}. The multiplier $A_{0}$ enforces the
Gauss constraint,
\be
\Omega = \pa_{i}\pi_{i} - \frac{m_{-}}{2}  \eps_{ij} \pa_{i} A_{j} \approx 0 
\label{an}
\ee 
where $(A_{i}, \pi^{i})$ is a canonical set. The hamiltonian on the constraint
surface is given by,
\be
H = \frac{1}{2} \int d^2x [ \pi_{i}^{2} + \frac{1}{2}F_{ij}^{2} + m_{-} \eps_{ij}\pi_{i} A_{j} + \frac{m_{-}^{2}}{4} A_{i}^{2}] \label{ao}
\ee 
Next, consider the canonical transformation,
\begin{eqnarray}
A_{i} & = & \frac{\pa_{i}}{\sqrt{-\pa^ {2}}} \pi_{\theta} + \eps_{ij}\frac
{\pa_{j}}{\sqrt{-\pa^ {2}}} \beta \nonumber \\
\pi_{i} & = & \frac{\pa_{i}}{\sqrt{-\pa^ {2}}} \theta - \eps_{ij}\frac{\pa_{j}}
{\sqrt{-\pa^ {2}}} \pi_{\beta} \label{ap} 
\end{eqnarray}
where $( \theta, \pi_{\theta})$ and $( \beta, \pi_{\beta})$ form independent
canonical pairs. Since this is a gauge theory, a gauge fixing is imposed.
We take the standard Coulomb gauge,
\be
\pa_{i}A_{i} = 0 \label{aq}
\ee
The presence of the gauge, together with the constraint (\ref{an}),  modifies
the canonical structure of the $(A_{i}, \pi_{i})$ fields; i.e. their brackets
are no longer canonical. The modified algebra can be obtained either by the 
Dirac algorithm \cite{PD} or, as done here, by just solving the constraints.
Their solution leads to the following structure,
\begin{eqnarray}
A_{i} & = & \eps_{ij}\frac{\pa_{j}}{\sqrt{-\pa^ {2}}} \beta \nonumber \\
\pi_{i} & = & - \frac{m_{-}}{2} \frac{\pa_{i}}{\sqrt{-\pa^ {2}}} \beta - 
\eps_{ij}\frac{\pa_{j}}{\sqrt{-\pa^ {2}}} \pi_{\beta} \label{ar}
\end{eqnarray}
which satisfies a nontrivial algebra,

\begin{eqnarray}
[ A_{i}(x), \pi_{j}(y)]  &=& i (-\delta_{ij}+\frac{\pa_{i}\pa_{j}}
{\pa^{2}} ) \delta(x-y) \nonumber\\ 
\left [ \pi_i(x),\pi_j(y) \right] &=& -i\frac{m_{-}}{2}\eps_{ij}\delta(x-y)
\label{as}
\end{eqnarray}

The same result follows by replacing the Poisson bracket by the Dirac bracket.
Using (\ref{ar}) the reduced hamiltonian is obtained from (\ref{ao}),
\be
H  =  \frac{1}{2}\int d^{2}x [(\pa_{i}\beta)^{2} + \pi_{\beta}^{2} + 
m_{-}^{2}\beta^{2}] \label{at} 
\ee
which has exactly the same structure as the second relation in (\ref{al}). 
Likewise the other Maxwell-Chern-Simons theory with a coupling $m_{+}$ can 
be reduced to the first relation in (\ref{al}). It might be mentioned
that the two lagrangians (\ref{l}) and (\ref{m}) differ not only in the
respective mass parameters, but also in the signature of the Chern-Simons
term. However a scaling argument shows that, apart from the field dependencies,
these are connected by $m_+ \rightarrow -m_-$. Since the hamiltonian is 
quadratic in the mass term, this sign difference 
therefore does not affect the result. 

Thus the reduced hamiltonian of the Maxwell-Chern-Simons theory with a mass 
term is the sum of the reduced hamiltonians of a doublet of Maxwell-Chern-
Simons theories with distinct mass parameters $m_{\pm}$. There is a complete
correspondence between the lagrangian and hamiltonian formulations.

\section{The energy momentum tensor and spin}
As emphasised in \cite{DJT}, spin in $2 + 1$ dimensions cannot be properly
identified from only the angular momentum operator since it does not conform 
to the conventional algebra. It is essential to consider the complete energy 
momentum tensor. Incidentally, although $\alpha$ and $\beta$ in (\ref{al})
satisfy the Klein-Gordon equation, these cannot be regarded as scalars due
to presence of the factor $ \sqrt{-\pa^{2}}$ in the transformations (\ref{ah}).
A complete analysis of the energy momentum tensor will be done which 
unambiguously determines the spin of the excitations.
The energy momentum tensor following from (\ref{ab}) is given by,
\begin{eqnarray}
\Theta_{\mu\nu} & = & 2 \frac{\pa {\cal L}}{\pa g^{\mu\nu}} - g_{\mu\nu} 
{\cal L} \nonumber \\
 & = & - F_{\mu\alpha}F_{\nu}^{\alpha} + m^{2} A_{\mu}A_{\nu} + \frac{1}{4}
g_{\mu\nu} F_{\alpha\beta}F^{\alpha\beta} - \frac{m^{2}}{2} g_{\mu\nu} 
A_{\alpha}A^{\alpha} \label{au}
\end{eqnarray}
The discussion of the hamiltonian has already been done. The momentum is given
by,
\begin{eqnarray}
P_{i} & = & \int d^{2}x \Theta_{0i} \nonumber \\
& = & \int d^{2}x ( - F_{0j}F_{i}^{j} + m^{2} A_{0}A_{i}) \label{av} 
\end{eqnarray}
To pass over to the reduced variables, $A_{0}$ is first eliminated by using
the constraint (\ref{af}). Next, the canonical transformations (\ref{ah})
and (\ref{ak}) are applied. This leads to the diagonal form,
\be
P_{i} = \int d^{2}x [ \pi_{\alpha}\pa_{i}\alpha + \pi_{\beta}\pa_{i}\beta ]
\label{aw}
\ee 
The rotation generator is given by,
\be
M_{ij} = \int d^{2}x [x_{i}\Theta_{0j} - x_{j}\Theta_{0i} ] \label{ax}
\ee
which, following the same techniques, is put in the diagonal form,
\be
M_{ij} = \int d^{2}x [ (x_{i} \pi_{\alpha} \pa_{j}\alpha  -  x_{j}
\pi_{\alpha} \pa_{i}\alpha) + (x_{i} \pi_{\beta} \pa_{j}\beta - x_{j}  
\pi_{\beta} \pa_{i}\beta ) ]\label{ay} 
\ee
Both the translation and rotation generators have their expected forms with
the fields $\alpha$ and $\beta$ transforming normally. Using the inverse
transformation of (\ref{ah}) it is seen that the original field $A_{i}$
also transforms normally,
\begin{eqnarray}
[A_{j}, P_{i}]  & = & i \pa_{i} A_{j} \nonumber \\
\left [A_{k}, M_{ij}\right ] & = & i (x_{i}\pa_{j}A_{k} - x_{j}\pa_{i}A_{k} - \delta
_{ik}A_{j} + \delta_{jk}A_{i}) \label{az} 
\end{eqnarray}
Finally, the boosts are considered and it is found that the diagonal form is 
given by,
\begin{eqnarray}
M_{0i} & = & t\int d^2x \Theta_{0i} - \int d^2x  x_{i}\Theta_{00} \nonumber \\  
 & = & t\int  d^2x \pi_{\alpha} \pa_{i}\alpha - \frac{1}{2} \int  d^2x  x_{i} [
(\pa_{j}\alpha)^{2} + \pi_{\alpha}^{2} + m_{+}^{2}\alpha^{2} ] + m_{+}
\eps_{ij} \int  d^2x \pi_{\alpha} (\frac{\pa_{j}}{\pa^{2}}) \alpha  \nonumber\\
& & + t\int  d^2x \pi_{\beta}\pa_{i}\beta - \frac{1}{2} \int  d^2x x_{i} [
(\pa_{j}\beta)^{2} + \pi_{\beta}^{2} + m_{-}^{2}\beta^{2} ] - m_{-}
\eps_{ij} \int  d^2x \pi_{\beta} (\frac{\pa_{j}}{\pa^{2}}) \beta \label{ba}
\end{eqnarray}
The boost generator has extra factors which clearly show that $\alpha$ and
 $\beta$ do not transform as scalars. These extra pieces are however essential
to correctly reproduce the usual transformation of the  original vector
field $A_{i}$,
\be
[A_{j}, M_{0i}] = i (t \pa_{i}A_{j} - x_{i}\pa_{0}A_{j} + \delta_{ij}A_{0})
\label{bb}
\ee
where recourse has to be taken to the solution of the constraint (\ref{af}) 
to obtain the final structure involving $A_{0}$.

The presence of the abnormal terms in the boost leads to a zero momentum
anomaly in the Poincare algebra,
\be
[ M_{0i}, M_{0j}]  = i( M_{ij} + \eps_{ij} \Delta \label{bc})
\ee
where,
\be
\Delta = \frac{m_{+}^{3}}{4\pi} \left (\int d^{2}x \alpha \right)^{2} + \frac{m_{+}}{4\pi}
\left(\int d^{2}x \pi_{\alpha}\right)^{2} -  \frac{m_{-}^{3}}{4\pi} \left(\int 
d^{2}x \beta \right)^{2} - \frac{m_{-}}{4\pi} \left(\int d^{2}x \pi_{\beta}
\right)^{2}
\ee
Following exactly the same steps as in \cite{DJT} it is possible to remove 
this anomaly, simultaneously fixing the spin of the excitations. Consider
the mode expansions,
\begin{eqnarray}
\alpha(x)& = & \int \frac{d^{2}k}{2\pi\sqrt{2\omega(k)}} [a(k) e^{-ik.x}
 + a^{\dagger}(k)e^{ik.x} ]  \nonumber\\
\beta(x) & = & \int \frac{d^{2}k}{2\pi\sqrt{2\omega(k)}}[b(k) e^{-ik.x}
 + b^{\dagger}(k)e^{ik.x} ] \label{be}
\end{eqnarray}
suitably modified by the phase redefinitions,
\be 
a \rightarrow e^{-i\phi}a, \hskip 1 cm b \rightarrow e^{ i\phi}b \label {bf} 
\ee
where,
\be
\phi  = \tan^{-1}\left(\frac{k_{2}}{k_{1}}\right)
\ee
It leads to the following expressions for the boosts and rotation generator,
\begin{eqnarray}
M_{0i} &=& \frac{i}{2} \int d^2k \omega (k) \vert {a}^\dagger(k)
\stackrel{\leftrightarrow}{\pa}_{i} a(k)\vert + \eps_{ij} \int d^2k
\frac {1}{\omega (k) + m_{+}} k_j {a}^\dagger(k)a(k) \nonumber\\
& & \frac{i}{2} \int d^2k \omega (k) \vert b^\dagger(k) \stackrel{\leftrightarrow }{\pa}_{i} b(k)\vert - \eps_{ij} \int d^2k\frac {1}{\omega (k) + m_{-}} k_j 
{b}^\dagger(k)b(k) \label {bg}
\end{eqnarray}

\begin{eqnarray}
M_{ij} & = &  \eps_{ij}\left(\int d^2k {a}^{\dagger}(k) \frac{1}{i} \frac{\pa}
{{\pa }{\phi}} a(k) - \int d^2k  {a}^{\dagger} (k)a(k)\right) \nonumber\\
&  &  +  \eps_{ij}\left(\int d^2k  b^{\dagger}(k) \frac{1}{i} \frac{\pa}{\pa 
\phi} b(k) + \int d^2k  b^{\dagger}(k)b(k) \right) \label {bg1}
\end{eqnarray} 
which satisfy the Poincare algebra 
\be
[M_{0i}, M_{0j}] = i M_{ij} \label {bh}
\ee

An inspection of the rotation generator shows that it comprises of two
distinct terms denoted by the parentheses. The first factor in each
corresponds to the usual orbital part. The additional pieces show that the spin of the excitations 
associated
with $\alpha$ and $\beta$ are, respectively, $-1$ and $+1$. 
This also happens in the case of the Maxwell-Chern-Simons theory \cite{DJT}.
The difference from
the spin of the excitations in the Maxwell-Chern-Simons theory is noteworthy.
There the sign of the spin is fixed by the sign of the coefficient of the Chern
-Simons parameter. In the present case it is seen from (\ref{am}) that, 
irrespective of the sign of $\theta$, the mass parameters $m_{\pm}$ are
always positive. Hence the sign of the spin associated with $\alpha$ and 
$\beta$ is also uniquely determined. 

Note that for $m_{+} = m_{-}$, the theory becomes parity conserving. 
This is the case when the Maxwell-Chern-Simons doublet with identical mass yields the Proca model \cite{D,BW}.

\section{Application to 3d bosonisation}
Bosonisation in higher dimensions is neither complete nor exact as in the case of 
two space-time dimensions. This is related to the fact that the fermion determinant
in dimensions greater than two cannot be exactly computed. In general it has a nonlocal
structure. However, for the large fermion mass limit in three space-time 
dimensions, a local expression emerges \cite{DJT,L}.
This has been exploited to discuss the bosonisation of massive fermionic models in the
long wavelength limit \cite{B}.  Here we analyse the bosonisation of a doublet 
of such models.
To be specific, consider the following three dimensional massive Thirring 
models,

\begin{eqnarray}
{\cal L_{+}}  &= & \bar{\psi} (i \pa {\llap {/{}}} + m_{+}) \psi - \frac{\lambda_{+}^{2}}
{2} (\bar{\psi}\gamma_{\mu}\psi)^{2}\nonumber\\ 
{\cal L_{-}} &= & \bar{\chi}(i \pa {\llap {/{}}} - m_{-})\chi - \frac{\lambda_{-}^{2}}{2}
(\bar{\chi}\gamma_{\mu}\chi)^{2} 
\label {bi}
\end{eqnarray}

The respective partition functions, after eliminating the four fermion interaction by introducing auxiliary fields, are given by,

\begin{eqnarray}
{\cal Z_{+}} &=& \int d\psi d\bar{\psi} df_{\mu} \exp i\int d^{3}x \left(\bar
{\psi} (i \pa {\llap {/{}}} + m_{+} + \lambda_{+} f{\llap{/{}}})\psi + \frac{1}
{2} f_{\mu}f^{\mu}\right )\nonumber \\
{\cal Z_{-}} &= &\int d\chi d\bar{\chi} dg_{\mu} \exp i\int d^{3}x\left(\bar
{\chi}(i \pa {\llap {/{}}} - m_{-}+ \lambda_{-} g{\llap{/{}}})\chi+ \frac{1}
{2} g_{\mu}g^{\mu} \right )\label {bj} 
\end{eqnarray}

The fermion determinant can be expressed, in the large mass limit, by a local
series involving $(\frac{\pa}{m})$ \cite{DJT,L,DR}. Furthermore, for weak coupling we need to 
consider only the two legs fermion loop. The leading long wavelength term
in this quadratic approximation is the Chern-Simons three form. Thus the 
effective bosonised lagrangians of the doublet are given by,

\begin{eqnarray}
{\cal L_{+}} &=& \frac{\lambda_{+}^{2}}{8\pi} \eps_{\mu\nu\lambda} f^{\mu}\pa^
{\nu} f^{\lambda} + \frac{1}{2} f_{\mu}f^{\mu} + O(\frac{1}{m})
\nonumber \\ 
{\cal L_{-}} &=& - \frac{\lambda_{-}^{2}}{8\pi} \eps_{\mu\nu\lambda} g^{\mu}\pa^
{\nu} g^{\lambda} + \frac{1}{2} g_{\mu} g^{\mu} + O(\frac{1}{m})
\label {bk}
\end{eqnarray}
\\
where the difference in the sign of the Chern-Simons piece is a result of a 
similar feature in the mass terms of the original lagrangians (\ref{bi}).

Using our previous results, the doublet of $ {\cal L_{+}}$ and $ {\cal L_{-}}$,
as defined in (\ref{bk}), can 
be represented by an effective
lagrangian, which is just the Maxwell-Chern-Simons
theory with an explicit mass term,

\be
{\cal L} =  - \frac{1}{4} F_{\mu\nu}F^{\mu\nu}
(A) + \frac{2\pi}{{\lambda_{+}^{2}}{\lambda_{-}^{2}}} (\lambda_{+}^{2} - \lambda_
{-}^{2}) \eps_{\mu\nu\lambda} A^{\mu}\pa^{\nu}A^{\lambda} + \frac{8\pi^{2}}
{{\lambda_{+}^{2}}{\lambda_{-}^{2}}}A_{\mu}A^{\mu} 
\label {bl}
\ee
where,
\be
A_{\mu} = \frac{\lambda_{+}\lambda_{-}}{\sqrt{4\pi(\lambda_{+}^{2} + \lambda_{-}
^{2})}} (f_{\mu} - g_{\mu}) \label {bm}
\ee
The lagrangian (\ref{bl}) can be regarded as a bosonised lagrangian
obtained from the following massive Thirring model,

\be
{\cal L} = \bar{\Psi}(i \pa {\llap{/{}}} - m)\Psi - \frac{\lambda^{2}}{2}
(\bar{\Psi} \gamma_{\mu}\Psi)^{2} 
\label {bn}
\ee
\\
in the weak coupling and large mass limit. The relations of the parameters
occurring in the above lagrangian and (\ref{bl}) are given by,

\begin{eqnarray}
m &=& \frac{8\pi}{3} \frac{(\lambda_{+}^{2} - \lambda_{-}^{2})}{\lambda_{+}^{2}
\lambda_{-}^{2}}\nonumber \\
\lambda^{2} &= & \lambda_{-}^{2} - \lambda_{+}^{2} 
\label {bo}
\end{eqnarray}

To show this we first observe that the original weak coupling involving 
$\lambda_{+}$ and $\lambda_{-}$ leads to a weak $\lambda$. Secondly, it also
implies the large mass limit. In other words the same approximation prevails.
The fermion determinant, similar to (\ref{bk}), but evaluated
to the next to leading order, includes both the Chern-Simons term  and 
the Maxwell term \cite{DR}. Specifically, this is written as,

\be
{\cal L} = \frac{1}{2}A^{\mu}A_{\mu} - \frac{\lambda^2}{8\pi}\eps{\mu
\nu\lambda}A^{\mu}\pa^{\nu}A^{\lambda} + \frac{\lambda^2}{24\pi m}
F^{\mu \nu}F_{\mu \nu} + O(\frac{1}{m^2})
\label{x}
\ee
\\
where we have identified the auxiliary field necessary to simplify the four
fermion interaction with
$A_{\mu}$. This is exactly in keeping with the spirit of obtaining (\ref{bk})
from (\ref{bj}), except that the fermion determinant has been evaluated to the
next to leading order in the inverse mass expansion. By making the following
scaling,

\be
A_{\mu}\rarrow \frac{4\pi}{\lambda_+\lambda_-}A_{\mu}
\label{y}
\ee
\\
this reproduces
(\ref{bl}), with the identification (\ref{bo}). 
This establishes the connection between (\ref{bi}) and (\ref{bn})
since (\ref{bl}) is their common origin. 

The implications of the above analysis are now discussed. 
In the  quadratic approximation, a doublet of massive Thirring
models in the leading long wavelength limit bosonises to the effective
lagrangian (\ref{bl}). The same effective theory, under similar
approximations and with the identification
(\ref{bo}), also characterises a single massive Thirring model, but where
the calculation of the fermion determinant is carried out till the first
non leading term.  In this sense, therefore, a doublet of massive Thirring
models can be approximated by  a single similar 
model. There is also a matching in the degree of freedom count.

\section{Conclusions}
We have considered the description of a doublet of self dual models with distinct
topological mass parameters, having opposite signs. The difference in sign 
implies that the doublet comprises a self dual and an anti-self dual model. 
Specifically, this was a pair of the gauge invariant Maxwell-Chern-Simons theory
\cite{DJT} or, equivalently, its dual gauge variant version \cite{TPN,DJ,BR}. 
The effective theory, 
characterising such a doublet, turned out to be the Maxwell-Chern-Simons theory with an 
explicit mass term. The basic field of the effective theory was just the 
difference of the doublet variables.

A canonical analysis of the effective theory was done. Based on a set of 
canonical transformations, the  hamiltonian was diagonalised into two separate
pieces. The two massive modes were found to be a combination of the topological
and explicit mass parameters. In fact these were identified with the two modes 
of the Maxwell-Chern-Simons doublet that led to the effective theory. In this 
way a correspondence was established between the  lagrangian approach of combining 
the doublet into an effective theory and the hamiltonian approach of decomposing 
the  latter back into the doublet. The spin of the excitations was obtained from
a complete study of the Poincare algebra by adopting the method advocated in
\cite{DJT}.  

When the Maxwell-Chern-Simons doublet has identical topological mass $\pm m$,
parity is conserved since one degree of freedom is just mapped to the other.
The spin carried by the two degrees of freedom is $\mp 1$. This has the same 
kinematical structure as the Proca theory which is a parity conserving theory 
with two massive modes having spin $\mp 1$ \cite{BI,DJT}. An explicit 
demonstration of this was provided earlier \cite{D,BW}. This result is 
reproduced here by putting $m_{+} = m_{-}$.
 
For the more general case where the Maxwell-Chern-Simons doublet has different
topological masses $ m_{\pm}$, parity is no longer conserved, although 
the other considerations remain valid. Hence the kinematics of such a doublet
resembles a non gauge parity violating theory with two massive modes having
spin $\mp 1$. This turned out to be the Maxwell-Chern-Simons theory with an 
explicit mass term, as elaborated here in details. An added bonus of this 
equivalence is that it led to fresh insights into the bosonisation of massive
fermionic models. This was explicitly shown for a doublet of massive Thirring
models, but it can be done for other examples like QED in three dimensions.

Recently there have been certain discussions \cite{KF,BM} which regard a mass
term in a gauge theory either as a conventional mass term or, equivalently,
as a gauge fixing term. In fact, Maxwell theory in the covariant gauge and
the Proca model were shown equivalent from the viewpoint of quantum BRST
symmetry \cite{KF,BM}. Here we find that the superposition of a pair of 
Maxwell-Chern-Simons theories in the covariant gauge leads to an explicit mass
generation. This suggests a possible connection between these different 
approaches.

The extension of these findings to higher dimensions or non abelian versions 
would be welcome. Of course for $4k-1$ dimensions where self duality is 
definable, this extension is straightforward in the abelian case. For non 
abelian theories, the superposition principle does not work as in the abelian 
theory. Using some special properties of two dimensions, the WZW non abelian 
doublet was treated in \cite{ABW}. But for general dimensions, the non abelian 
analysis remains an open issue.


\end{document}